\documentclass{ws-procs9x6}            
\usepackage{amsfonts}
\usepackage{subfigure}
\usepackage{amsmath}
\usepackage{epsfig}
\usepackage{bm}
\usepackage[utf8]{inputenc}

\begin{document}

\title{Quantum  heat engines  with  multiferroic working substance
\footnote{This work is supported by the DFG through SFB 762.}}

\author{L. Chotorlishvili$^{\dag }$, M. Azimi, S. Stagraczy\'nski, J. Berakdar}
\address{Institut f\"ur Physik, Martin-Luther-Universit\"at Halle-Wittenberg, \\
06099 Halle, Germany\\
$^\dag $E-mail: levan.chotorlishvili@physik.uni-halle.de }


\begin{abstract}
The work provides an overview on some recent advances in the area of  quantum thermodynamics and quantum heat engines. A particular emphasis is  put on the possibility of constructing finite time quantum cycles and adiabatic shortcuts.
We discuss in details the particular quantum heat engines operating  with a multiferroic working substance.
\end{abstract}


\bodymatter

\section{ Introduction}

For any thermal heat engines and thermodynamic cycles the three main criteria are central: The produced maximal work, the efficiency, and the output power of the engine.
The high efficiency of the heat engine is  important for performing   operations with low energy consumption,
 while the amount of the produced work and the power of the heat machine are crucial for swift  performance. Is it realistic  to meet all those three criteria simultaneously?.
 The present work  addresses this question and provides a brief overview on the current status of knowledge.\\
 To introduce the problem and notations we start by recalling   the basics of the classical thermodynamics \cite{Landau5}.
For a thermally isolated system consisting of  two parts which are not in thermal equilibrium, a certain work is performed on the surrounding bodies during the transition to the equilibrium state. We exclude the work associated with the general expansion, assuming that the total volume of the system is conserved. Then the produced work is a function of the internal energy of the system $|W|=U_{0}-U(S)$ where $U_{0}$ is the initial energy. Since the transition to the equilibrium state might occur in a variety  of ways the final energy and the entropy $U(S)$ might be different. Considering the produced work as a functional of the system's entropy we write $\delta_S|W|=-\big(\partial U(S)/\partial S\big)_{V=const}=-T$.  As $\delta_S |W|$ is always negative (we use the absolute temperature scale $T>0$),
 any increment in the entropy while producing the work lowers work. Hence, the  maximal work $|W|_{max}$ is produced during the process when  the
 entropy of the system stays constant $S=const$. More rigorous deliberation leads to the following formula for the
 maximal work $|W|_{max}=-\delta \big(U-T_{0}S+P_{0}V\big)$. Here $T_{0},P_{0}$ stand for the temperature and the pressure of the environment, while $U,S,V$ define the internal energy, the entropy,  and the volume of the working body.
 If the volume and the temperature  are constant during the process, the produced maximal work is equal to the change in the free energy $|W|_{max}=-\delta F$.
  Note, the adiabaticity of the process excludes any direct energy exchange between the bodies with the different temperatures.

The ideal heat engine envisioned  by Carnot has four strokes:  The working substance at temperature $T_{h}$  absorbs isothermally
energy from the hot heat bath, and then  is  cooled down adiabatically to the temperature $T_{l}$.
Thereafter, the  working substance  releases isothermally energy to a cold heat bath at $T_{l}$,  and eventually returns   adiabatically  to the initial state.
  Two thermal baths with temperatures $T_{h}>T_{c}$ are needed  to perform a reversible cycle.
  The existence of two heat baths allows excluding a  direct irreversible energy exchange between the systems with different temperatures.
  The efficiency and the produced work of the ideal heat engine read:
  $\eta=\frac{T_{h}-T_{l}}{T_{h}}$, $W_{max}=\frac{T_{h}-T_{l}}{T_{h}}Q_{in}$,
  where $Q_{in}$ is the heat absorbed from the hot bath.
   The adiabaticity imposes certain restrictions on the cycle's swiftness.
   An ideal cycle takes an infinite time $\tau\mapsto\infty$, and
therefore the output power of the ideal Carnot cycle vanishes  $P=W_{max}/\tau =0$.
So it is of relevance to find
ways  yielding a good cycle efficiency with a reasonable power.
 This issue  is not only a technical but also it is  a  conceptual one,  even for classical systems.
 For quantum heat engines additional aspects are important.
 When the size of the working medium is scaled down to the mesoscopic scale purely quantum effects such as quantum fluctuations and interlevel transitions become important.
 The problem of the thermally assisted interlevel transitions can be solved relatively easily by detaching the
 working body from the heat bath.\\
Quantum adiabaticity is more subtle.
 To be precise we specify the concept of adiabaticity separately for quantum and classical systems.
 The stroke of the cycle, which is adiabatic in the sense of classical thermodynamics may be nonadiabatic for a quantum working substance.
 The  reason lies in  quantum interlevel transitions that naturaly occur in the case of fast driving and a  finite time thermodynamic cycles.
 Thus, quantum adiabaticity implies not only a decoupling of the system from the thermal source, but also requires an elimination of interlevel transitions
 that are of a pure quantum nature. Shortcuts to quantum adiabaticity is a recent theoretical concept that allows eliminating the effect of those interlevel transitions \cite{Demirplak2003,Demirplak2005,Torrontegui,Berry,delCampo2012,delCampo2014,Beau2016,Chotorlishvili}. In what follows we provide a brief introduction to
  shortcuts to quantum adiabaticity. Before that we discuss the
  efficiency of the Carnot engine at maximum output power for a classical system \cite{Curzon}.

\section{Efficiency of the Carnot engine at  a maximum output power}
An ideal Carnot engine assumes that during isothermal strokes the working substance is in equilibrium with the thermal reservoirs, meaning that the
 isothermal strokes are performed infinitely slowly. Therefore, the
power output of the engine is zero, since the finite work is produced in an infinite time.
Ref.\cite{Curzon} assumes that during the isothermal expansion the
 heat flux through the vessel enfolding the working medium is proportional to the temperature gradient across the vessel. Therefore,  the
 expressions for the absorbed heat and the heat rejected to the heat sink read
  $Q_{in}=\alpha t_{1}\big(T_{h}-T_{hw}\big), Q_{out}=\beta t_{1}\big(T_{lw}-T_{l}\big)$. Here $\alpha,\beta$ are constants, $t_{1},t_{2}$  are the durations
   of the isothermal strokes, and $T_{hw},T_{lw}$ are the temperatures of the working medium during the isothermal strokes. The key issue in this assumption is that the duration of the
isothermal strokes are finite.  However,
 the cycle is reversible and the total entropy production  is zero $Q_{in}/T_{hw}=Q_{out}/T_{lw}$. The output power of the cycle is $P=\big(Q_{in}-Q_{out}\big)/\big(t_{1}+t_{2}\big)$ and the total time spent for the two adiabatic strokes is $\big(\gamma-1\big)\big(t_{1}+t_{2}\big)$.  With
 this expressions one can maximize the power output $\partial P/\partial x=0, \partial P/\partial y=0$, where $x=T_{h}-T_{hw}, y=T_{lw}-T_{l}$. 
 After a little algebra for the cycle efficiency  and the maximum output
power one obtains: $\acute{\eta}=1-\big(T_{l}/T_{h}\big)^{1/2}$. Note a finite time cycle needed for the  maximal power output comes at the cost of   a
 lower efficiency $\acute{\eta}<\eta$. Unfortunately, the  results obtained in Ref.\cite{Curzon} are not directly applicable to quantum heat engines since in the quantum case pure quantum interlevel transitions may lift the adiabaticity.   Adiabatic shortcuts are thus needed.

\section{First law of thermodynamics for quantum systems}

The first law of thermodynamics states  that the change in internal energy of a system is equal to the heat added to the system minus the work done by the system. This is a very general formulation
applicable to  quantum systems as well. However, for  quantum systems the definitions of a quantum heat and a quantum work need to be revisited\cite{Quan2006,Quan2007,Quan2009,Esposito,Kumar,Alecce,Abah2012,Abah2014,Rossnagel,WangR,Cakmak,Zagoskin,Altintas,Ivanchenko,Linden,Broeck,Azimi2,Mukherjee,Kosloff,AbahLutz}.

At nonzero temperature  the energy of a system can be evaluated as follows $U=Tr\big(\hat{\varrho}\hat{H}\big)$. Here $\hat{H}$ is the Hamiltonian of the system and $\hat{\varrho}$ is the density matrix. For the change
in the internal energy we deduce: $dU=\sum_{n=1}^{N}\big(E_{n}d\varrho_{nn}+\varrho_{nn}dE_{n}\big)$. The first term $\delta Q= E_{n}d\varrho_{nn}$  corresponds to the heat exchange and is related to the change of the level populations $\varrho_{nn}\big(E_{n},T)$ occurring due to a change in the temperature $T$ for $E_{n}=const$. The second term $\delta W=\varrho_{nn}dE_{n}$ corresponds to the produced work. The working substance produces work due to a change in the energy spectrum $dE_{n}$ of the system.  The relation of  heat exchange and the quantum level populations is clear. The concept of a quantum work needs however further specification \cite{Talkner,Bochkov1,Bochkov2,Bochkov3,Bochkov4,Jarzynski1,Pitaevskii,Hermans,Deffner}.

\section{Bochkov-Kuzovlev and Jarzynski equalities}

Let us consider a thermally isolated classical system $H\big(p,q,\lambda(t)\big)$ driven by an external parameter $\lambda$. A change in the parameter $\lambda(0)=\lambda_{0}, \lambda(t_{f})=\lambda_{f}$  produces work delivered to the system. This work is assumed to be small compared to the energy of the system. At $t=0$ the
system is thermalized to a temperature $T=1/\beta$. The work done on the system reads\cite{Talkner} $$W=\int_{0}^{t_{f}}\frac{\partial H}{\partial\lambda}\frac{d\lambda}{dt}dt=H\big(p_{f},q_{f},\lambda(t_{f})\big)-H\big(p_{0},q_{0},\lambda(0)\big).$$
The work averaged over the statistical ensemble is $$\big<W\big>=\big<H\big(p_{f},q_{f},\lambda(t_{f})\big)\big>-\big<H\big(p_{0},q_{0},\lambda(0)\big)\big>.$$
 According to the Bochkov-Kuzovlev equalities,  not the work $\big<W\big>$ itself, but the exponential of the work $\big<\exp(-\beta W)\big>$  is the key point: $$\big<\exp(-\beta W)\big>=\int\exp\big[\beta(F_{0}-H_{0})\big]\exp\big[-\beta(H_{f}-H_{0})\big]d\Gamma_{0}.$$
  Here $\exp\big[\beta(F_{0}-H_{0})\big]$ is the distribution function of the equilibrated system at $t=0$, $F_{0}$ is the free energy, and $\Gamma_{0}=p_{0}q_{0}$ is the phase volume of the system. Due to the Liouville theorem the phase volume of the system is invariant $d\Gamma_{0}=d\Gamma_{f}$. Therefore, one writes
  $$\big<\exp(-\beta W)\big>=\exp(\beta F_{0})\int\exp(-\beta H_{f})d\Gamma_{f}.$$
   Finally we obtain the expression $\big<\exp(-\beta W)\big>=\exp(-\beta \Delta F)$, where $\Delta F=F_{f}-F_{0}$. This equality for cyclic process $\lambda_{f}=\lambda$ was obtained by Bochkov and Kuzovlev in 1977 and  by Jarzynski in 1997 for a more general setting  $\lambda_{f}\neq\lambda$.

According to statistical physics, for an adiabatic process when the parameter $\lambda$  changes slowly compared to the system's relaxation time, the
produced work is equal to the change in free energy $W=\Delta F$. For the non-adiabatic case $W>\Delta F$ part of the work is wasted on the entropy production. The nonequilibrium entropy associated with the nonadiabatic process is defined as $\Delta S_{ir}=\beta \big<W_{ir}\big>$. Here
$\big<W_{ir}\big>=\big<W\big>-\Delta F$ is the difference between the delivered work in the nonadiabatic process and the change in free energy.
We can rewrite Jarzynski equality in the following form (cf. Ref.\cite{Hermans}): $\Delta F=-T\log\big<\exp\big(-\beta W \big)\big>$. Using the
ansatz for the work fluctuations $W=\big<W\big>+\delta W$ we obtain  $\Delta F = \big<W\big> -\beta\big<(\delta W)^{2}\big>/2$.
Noteably, the work for a quantum system  is not an observable. This means that the average of the total work performed on the system doesn't corresponds to the expectation values of an operator representing the work \cite{Talkner}. In particular the work delivered to a quantum system is related to the time ordered correlation functions of the exponentiated Hamiltonian. The nonequilibrium entropy associated with a nonadiabatic process can be calculated straightforwardly \cite{Deffner}. First we define the probability distribution of the quantum work $$p\big(W\big)=\sum_{n,m}\delta\big(W-\big(E_{m}^{n}-E_{n}^{0}\big)\big)p_{n,m}^{\tau}p_{n}^{0}. $$
 Here $$p_{n}^{0}=\exp\big(-\beta E_{n}^{0}\big)/Z_{0}$$ is the initial thermal Gibbs distribution, $Z_{0}$ is the partition function,
$E_{n}^{0}$ are the initial energies, and $p_{n,m}^{\tau}$ are  the transition probabilities. Using $p\big(W\big)$ one can calculate the non-equilibrium quantum work:
$$\big<W\big>=1/\beta \sum_{n}p_{n}^{0}\ln p_{n}^{0}-1/\beta \sum_{n,m}p_{n}^{0}p_{n,m}^{\tau}\ln p_{n}^{\tau}-\Delta F.$$
  $p_{n}^{\tau}$ is the final equilibrium distribution function. In the absence of  purely quantum inter-level transitions $p_{n,m}^{\tau}=\delta_{nm}$ the first two terms disappear and the quantum work becomes equal to the change in free energy.



\section{Adiabatic shortcuts and finite time quantum cycles}

For a  general discussion of  shortcuts to adiabaticity and an overview of the interrelation between the various
approaches as well as  their historical developments  we refer to the review article \cite{Torrontegui} and references therein.  Here we will basically  follow
Berry's transitionless driving formulation\cite{Berry}  which  is equivalent to the counterdiabatic approach of Demirplak and Rice \cite{Demirplak2003,Demirplak2005}.

Let us suppose that the Hamiltonian of the system has the form $\hat{H}\big(p,q,\lambda \big)$. Here $p,q$ are canonical coordinates and $\lambda$ is a parameter in the sense discussed above. For the solution of Schr\"odinger equation $$\imath \frac{\partial \Psi}{\partial t}=\hat{H}\Psi$$ we implement the
 following ansatz: $$\Psi =\sum_{n}a_{n}\big(t\big)\varphi_{n}\big(p,q,\lambda\big)\exp \big\{-\imath \int_{-\infty}^{t}E_{n}\big(\lambda\big)dt\big\},$$
  where $E_{n}\big(\lambda\big)$ are the instantaneous quasi-energies that depend adiabatically  on the parameter $\lambda$. After standard derivations for the time dependent coefficients $a_{n}\big(t\big)$ we obtain the iterative solution $$a_{n}^{(1)}\big(t\big)=-\int_{-\infty}^{t}d\tau\sum_{m\neq n}\frac{\big\langle\varphi_{n}\big|\frac{\partial H}{\partial \lambda}\big|\varphi_{m}\big\rangle\dot{\lambda}}{E_{m}-E_{n}}\times a_{m}\big(-\infty\big)\exp \big\{-\imath \int_{-\infty}^{\tau}\big(E_{m}-E_{n}\big)d\acute{\tau}\big\}.$$
    The adiabatic approximation is valid when the following criteria hold
    $$\frac{a_{n}^{(2)}}{a_{n}^{(1)}}\sim \frac{\partial H}{\partial t}\frac{1}{\big(E_{m}-E_{n}\big)^{2}}. $$
    Here $a_{n}^{(2)}$ is a second order correction to $a_{n}\big(t\big)$. If the characteristic time scale of the parameter $\lambda$ is $\dot{\lambda}\sim 1/\tau$ and $\frac{\tau}{\big(E_{m}-E_{n}\big)^{2}}\gg 1$ then the dynamic of the system is adiabatic and the effect of the non-adiabaticity is exponentially small.

In the case of an adiabatic evolution the general   state $\vert\Psi_{n}(t)\rangle$ driven by $\hat{H}_{0}(t)$ is cast as
\begin{eqnarray}
\label{states}
&&\vert\Psi_{n}(t)\rangle=\exp\bigg[-\frac{i}{\hbar}\int_{0}^{t}dt^\prime E(t^\prime)\nonumber\\
&&~~~~~~~~~~~~~~~~-\int_{0}^{t}dt^\prime\langle\Phi_{n}(t^\prime )\vert\partial_{t^\prime}\Phi_{n}(t^\prime )\rangle\bigg]\vert\Phi_{n}(t)\rangle.
\end{eqnarray}
In  essence  the method of adiabatic shortcuts is an inverse engineering problem with the aim of finding of a new Hamiltonian for which the states (\ref{states}) behave as
$\imath\partial_{t}\Psi_{n}\big(t\big)=\hat{H(t)}\Psi_{n}\big(t\big)$. Note that the time dependence of the new Hamiltonian can be arbitrary fast.
With the aid of the unitary time-evolution operator
\begin{eqnarray}
\label{Unitary Operator}
&&\hat{U}(t)=\displaystyle\sum_{n}\exp\bigg[-\frac{i}{\hbar}\int_{0}^{t}dt^\prime E(t^\prime)\nonumber\\
&&~~~~~~~-\int_{0}^{t}dt^\prime\langle\Phi_{n}(t^\prime )\vert\partial_{t^\prime}\Phi_{n}(t^\prime )\rangle\bigg]\vert\Phi_{n}(t)\rangle\langle\Phi_{n}(0)|,
\end{eqnarray}
we construct the auxiliary (counter-diabatic)  Hamiltonian
\begin{eqnarray}
\label{auxiliary Hamiltonian}
\hat{H}_{CD}(t)=i\hbar\big(\partial_{t}\hat{U}(t)\big)\hat{U}^\dag(t).
\end{eqnarray}
The reverse state engineering  relies on the requirement that  the states (\ref{states}) solve for the Hamiltonian (\ref{auxiliary Hamiltonian}), meaning that
\begin{eqnarray}
\label{exact solving states}
i\hbar\partial_{t}\vert\Psi_{n}(t)\rangle=\hat{H}_{CD}(t)\vert\Psi_{n}(t)\rangle.
\end{eqnarray}
In this way even for a  fast driving  the transitions between the eigenstates $\vert\Phi_{n}(t)\rangle$ are prevented.
 After a  relatively simple algebra the counter-diabatic  (CD) Hamiltonian $\hat{H}_{CD}(t)$ takes the form
\begin{eqnarray}
\hat{H}_{CD}(t)=\hat{H}_{0}(t)+\hat{H}_{1}(t),
\end{eqnarray}
where
\begin{eqnarray}
\label{Hamiltonian1}
\hat{H}_{1}(t)=i\hbar\displaystyle\sum_{m\neq n}\frac{\vert\Phi_{m}\rangle\langle\Phi_{m}\vert\partial_{t}\hat{H}_{0}(t)\vert\Phi_{n}\rangle\langle\Phi_{n}\vert}{E_n - E_m}.
\end{eqnarray}
We adopt the initial conditions for the driving protocol as $$\hat{H}_{CD}(0)=\hat{H}_{0}(0), \quad \hat{H}_{CD}(\tau)=\hat{H}_{0}(\tau). $$  Thus,
on the time interval $t\in [0, \tau]$ we achieve a fast adiabatic dynamic by means of the counter-diabatic Hamiltonian $\hat{H}_{CD}(t)$.

This result can be straightforwardly generalized to  systems with a degenerated spectrum \cite{XueKeSong}. In this case we have
\begin{eqnarray}
\label{Hamiltonian2}
\hat{H}_{1}(t)=i\hbar\displaystyle\sum_{m\neq n}\sum_{q=1}^{\lambda_{n}}\sum_{k=1}^{\lambda_{m}}\frac{\vert\Phi_{m}^{k}\rangle\langle\Phi_{m}^{k}\vert\partial_{t}\hat{H}_{0}(t)\vert\Phi_{n}^{q}\rangle\langle\Phi_{n}^{q}\vert}{E_n - E_m}.
\end{eqnarray}
Here we assumed that the eigenvalue $E_{m}$ is $\lambda_{m}$ times degenerated and $\vert\Phi_{m}^{k}\rangle$ are the corresponding degenerated eigenfunctions $k=1,...\lambda_{m}$.

\section{Superadiabatic quantum heat engine with a multiferroic working medium}

\begin{figure}
	\centering
		\includegraphics[width=0.80\textwidth]{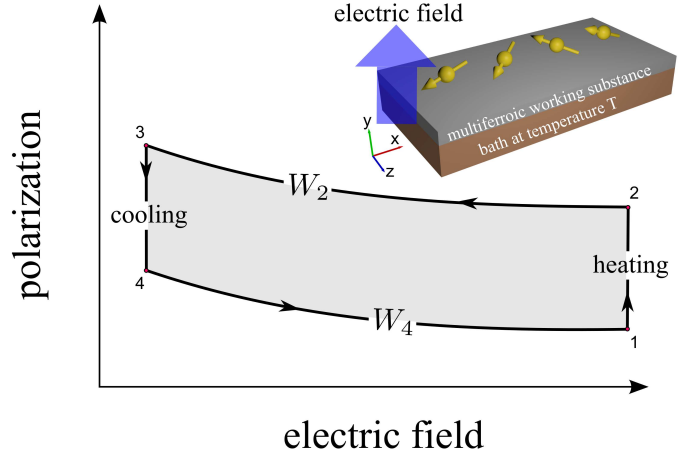}
	\caption{Scheme of the quantum Otto cycle based on a chiral multiferroic chain. Adapted from \cite{Chotorlishvili}}
	\label{fig:Scheme}
\end{figure}

A central point for any quantum heat engine is the choice of the appropriate working substance.
We identified  multiferroics (MF) and in particular magnetoelectrics nanostructures as promising candidates\cite{Chotorlishvili,Azimi2}.
MFs are materials of herostructures with    coupled order parameters such as elastic, magnetic, and ferroelectric ordering \cite{Eerenstein,WangJ,Garcia,Bibes,Spaldin,Valencia,Khomskii,Meyerheim,Cheong,Horley,Menzel,Wangkf} and can be well integrated in solid-state
electronic circuites (in particular in oxide-based electronics). Hence, an engine based on a MF substance performs  magnetic, electric and possibly (via piezoelectricity) mechanical works, at the same time.
Particularly relevant are   quantum spiral magnetoelectric substances \cite{Azimi1}. A prototypical  one dimensional chiral MF system can be modeled by a frustrated spin = 1/2 chain of $N$ sites aligned along the $x$ axis. Spin frustrations is due to competing ferromagnetic nearest neighbor $J_1>0$ and antiferromagnetic next-nearest neighbor $J_2<0$ interactions. We apply a time dependent electric field  $\wp(t)$ which is linearly polarized along the $y$ axis, and an external magnetic field $B$ is applied along the $z$  axis. The corresponding Hamiltonian reads
\begin{eqnarray}
\label{Hamiltonian0}
&&\hat{H}_{0}(t)=\hat{H}_{S}+\hat{H}_{SF}(t),\\
&&\hat{H}_{S}=-J_1\displaystyle\sum_{i}\vec{\sigma}_i\cdot\vec{\sigma}_{i+1}-J_2\displaystyle\sum_{i}\vec{\sigma}_i\cdot\vec{\sigma}_{i+2}-\gamma_{e}\hbar B\displaystyle\sum_{i}\sigma_{i}^{z},\nonumber
\end{eqnarray}
Here $\hat{H}_{S}$ is  time independent, while $\hat{H}_{SF}$ is  time dependent and contains the coupling of the external electric field to the electric polarization of the chain. The electric polarization  $\vec{P}_{i}$ tagged
to spin  non-collinearity  reads $$\vec{P}_{i}=g_{ME}\vec{e}_{i,\,i+1}\times (\vec{\sigma}_{i}\times\vec{\sigma}_{i+1}), $$ where $\vec{e}_{i,\,i+1}$ is the unit vector connecting the sites $i$ and $i+1$, $g_{ME}$ is a magneto-electric coupling constant. The spatially homogeneous, time dependent electric  field  $\wp(t)$ couples to the chain electric polarization $\vec{P}$ such that $$\vec{\wp}(t)\cdot\vec{P}=d(t)\displaystyle\sum_{i}(\vec{\sigma}_{i}\times\vec{\sigma}_{i+1})^z,$$
 with $d(t)=\wp(t)g_{ME}$. The quantity $(\vec{\sigma}_{i}\times\vec{\sigma}_{i+1})^z$ is known as the $z$ component of the vector chirality. With this notation $\hat{H}_{SF}(t)$ reads
\begin{eqnarray}
\label{Hamiltonian01}
&&\hat{H}_{SF}(t)=-\vec{\wp}(t)\cdot\vec{P}=d(t)\displaystyle\sum_{i}(\sigma_{i}^{x}\sigma_{i+1}^{y}-\sigma_{i}^{y}\sigma_{i+1}^{x}).
\end{eqnarray}
The quantum Otto cycle consists of two quantum isochoric and two adiabatic strokes. The quantum isochoric strokes correspond to a heat exchange between the working substance and the cold and the hot heat baths. During the quantum isochoric strokes the level populations are altered, see Fig.~\ref{fig:Scheme}.

The  MF working substance produces work during the adiabatic process.  Changing  the amplitude of the applied external electric field modifies the energy spectrum of the system. This is the mechanism behind producing work. The quantum Otto cycle and the  MF-based engine is detailed in recent works \cite{Chotorlishvili,Azimi2}. A particular type of the time dependence for the external electric field is given by
\begin{eqnarray}
\label{protocol}
d(t)=\epsilon\bigg(\frac{t^3}{3\tau}-\frac{t^2}{2}\bigg)+d_0,
\end{eqnarray}
which ensures that the requirement for the shortcuts of adiabaticity  $$\hat{H}_{CD}(0)=\hat{H}_{0}(0),\: \hat{H}_{CD}(\tau)=\hat{H}_{0}(\tau),$$ is fulfilled.
The power output of the quantum Otto cycle is given by
\begin{eqnarray}
\Re=\frac{-\big(\langle W_2\rangle_\mathrm{ad}+\langle W_4\rangle_\mathrm{ad}\big)}{\tau_1(T_H)+\tau_2+\tau_3(T_L)+\tau_4}.
\end{eqnarray}
Here $\tau_1(T_H)$, $\tau_3(T_L)$ are the relaxation times of the MF working substance in contact with the hot and the cold thermal baths (strokes $1\rightarrow 2$, and $3\rightarrow 4$), $\tau_2$ and $\tau_4$ correspond to the duration of the adiabatic strokes, $\langle W_2\rangle_\mathrm{ad}$ and $\langle W_4\rangle_\mathrm{ad}$ correspond to the work produced during the quantum adiabatic strokes. The condition
\begin{eqnarray}
&&\langle W_2\rangle_\mathrm{ad}+\langle W_4\rangle_\mathrm{ad}+Q_\mathrm{in}+Q_\mathrm{out}=0,\nonumber
\end{eqnarray}
during the whole cycle should be satisfied. The corresponding absorbed heat $Q_\mathrm{in}$ and the released heat $Q_\mathrm{out}$ by the working substance are defined as follows
\begin{eqnarray}
Q_\mathrm{in}&=& \sum_{n}E_{n}(0)\bigg(\frac{e^{-\beta_{H}E_n(0)}}{\sum_{n}e^{-\beta_H E_n(0)}}-\frac{e^{-\beta_{L}E_n(\tau)}}{\sum_{n}e^{-\beta_L E_n(\tau)}}\bigg),\nonumber\\
Q_\mathrm{out}&=&\sum_{n}E_{n}(\tau)\bigg(\frac{e^{-\beta_{L}E_n(\tau)}}{\sum_{n}e^{-\beta_L E_n(\tau)}}-\frac{e^{-\beta_{H}E_n(0)}}{\sum_{n}e^{-\beta_H E_n(0)}}\bigg).\nonumber\\
\label{heats}\end{eqnarray}

We adopt the dimensionless parameters $$J_1=1,\, J_2=-1,\, B=0.1,\, d_0=2.5,\, \epsilon=1.$$ In explicite units these parameters correspond to the one phase MF material\cite{Park}  $\textrm{LiCu}_{2}\textrm{O}_{2}$,   $J_1=-J_2=44[\mathrm{K}]$. The external driving fields strengths are $B=3[\mathrm{T}]$, $\wp=5\times10^{3}[\mathrm{kV/cm}]$.
We assume that the duration of the adiabatic strokes of the cycle are equal to $\tau_2=\tau_4=\tau$. The time unit in our calculations corresponds to the $\hbar/J_{1}\approx 0.1[\mathrm{ps}]$. CD driving allows reducing  the driving time. Implementing a short driving protocol is supposed  to maximize the output power of the cycle. Duration of the isochoric strokes can be estimated via the Lindblad master equation \cite{Breuer}.

We supplement the CD Hamiltonian $\hat{H}_{CD}(t)$ by the Hamiltonian of the heat bath $\hat{H}_\mathrm{bath}$ and the
 system-bath interaction $\hat{H}_\mathrm{int}$.
In addition,  we assume that the phononic heat bath is coupled to the $z$ component of the vector chirality $K_n^z=(\sigma_n^x\sigma_{n+1}^y-\sigma_n^y\sigma_{n+1}^x)$. The argument behind this doing  is that the vector chirality is a characteristic measure for the  non-collinearity in the spin order and is directly influenced by the lattice distortion and the phononic modes
\begin{eqnarray}
\label{interactions}
&&\hat{H}=\hat{H}_{CD}(t)+\hat{H}_\mathrm{int}+\hat{H}_\mathrm{bath},\nonumber\\
&&\hat{H}_\mathrm{bath}=\int dk \omega_k \hat{b}^{\dag}_{k}\hat{b}_{k},\nonumber\\
&&\hat{H}_\mathrm{int}=\displaystyle\sum_{n=1}^4 K_n^z\int dk g_k(\hat{b}^{\dag}_{k}+\hat{b}_{k}).
\end{eqnarray}
Here $\hat{b}^{\dag}_{k},~~\hat{b}_{k}$ are the phonon creation and annihilation operators, and $g_k$ is the coupling constant between the system and the bath. After a straightforward derivations we obtain
\begin{eqnarray}
\label{master equation}
&&\frac{d\rho_S(t)}{dt}=\displaystyle\sum_{\omega,\omega^\prime}\displaystyle\sum_{\alpha,\gamma}e^{i(\omega-\omega^\prime)t}\Gamma(\omega)
\big(K_{\beta}^{z}(\omega)\rho_S(t)K_{\alpha}^{z^\dag}(\omega^\prime)\nonumber\\
&&~~~~~~~~~~~~~~~~~~~~~-K_{\alpha}^{z^\dag}(\omega^\prime)\big(K_{\beta}^{z}(\omega)\rho_S(t)\big)+h.c.,\nonumber\\
&&\Gamma(\omega)=\int_{0}^{\infty}ds e^{i\omega s}\langle B^{\dag}(t)B(t-s)\rangle.
\end{eqnarray}
Here $$B(t)=\int dk g_k(\hat{b}^{\dag}_{k}e^{i\omega_{k}t}+\hat{b}_{k}e^{-i\omega_{k}t}), \; K_{\alpha}^{z}(\omega)=\displaystyle\sum_{\omega=E_m-E_n}\pi(E_n)K_{\alpha}^{z}\pi(E_n)$$ and
$\pi(E_n)=\vert\Psi_n\rangle\langle\Psi_n\vert$ is the projection operator onto the eigenstates $\vert\Psi_n\rangle$ of the CD Hamiltonian. For the bath correlation functions $\Gamma(\omega)$ we deduce
\begin{eqnarray}
\label{bathcorrelation}
&&\gamma(\omega)=\Gamma(\omega)+\Gamma^{\ast}(\omega),\nonumber\\
&&\gamma(\omega)=\pi J\big(\omega\big)\begin{cases} \frac{1}{\exp[\beta\omega]-1}, &  \omega< 0 \\
\frac{1}{\exp[\beta\omega]-1}+1, & \omega>0  \end{cases}.
\end{eqnarray}
Here $J\big(\omega\big)=\frac{\pi}{\omega}\displaystyle\sum_{j}g_{j}^{2}\delta(\omega-\omega_{j})=\pi\gamma$ is the spectral density of the thermal bath \cite{Breuer}.
The efficiency of the cycle reads $\eta = \frac{\delta Q_{H}+\delta Q_{c}}{\delta Q_{H}}$.

\section{Summary}

\begin{figure}[h]
\includegraphics[width=0.8\textwidth]{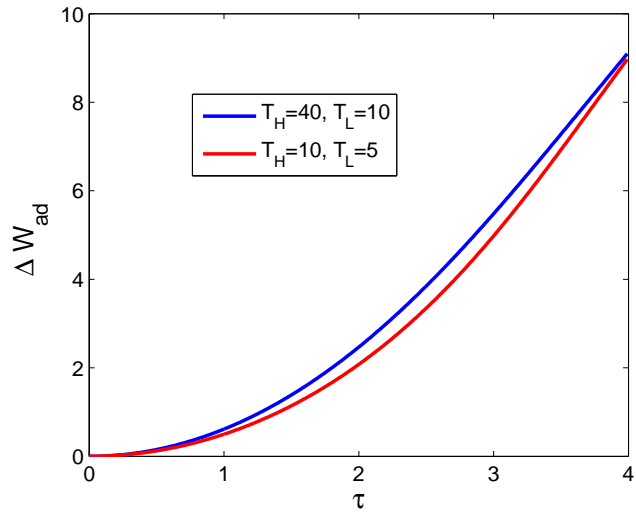}
\caption{\label{rms} Standard deviation of the work $\Delta W_\mathrm{ad}$ in scaled units for two different hot and cold bath temperatures.
 The other parameters are: $\varepsilon =1$, $J_1=1, J_2=-1, B=0.1, d_0=2.5$. Unscaled unit of $\Delta W_{ad}$ amounts  to $6\times10^{-22}[J]$. Adapted from \cite{Chotorlishvili}.}
\end{figure}
\begin{figure}
	\centering
		\includegraphics[width=0.80\textwidth]{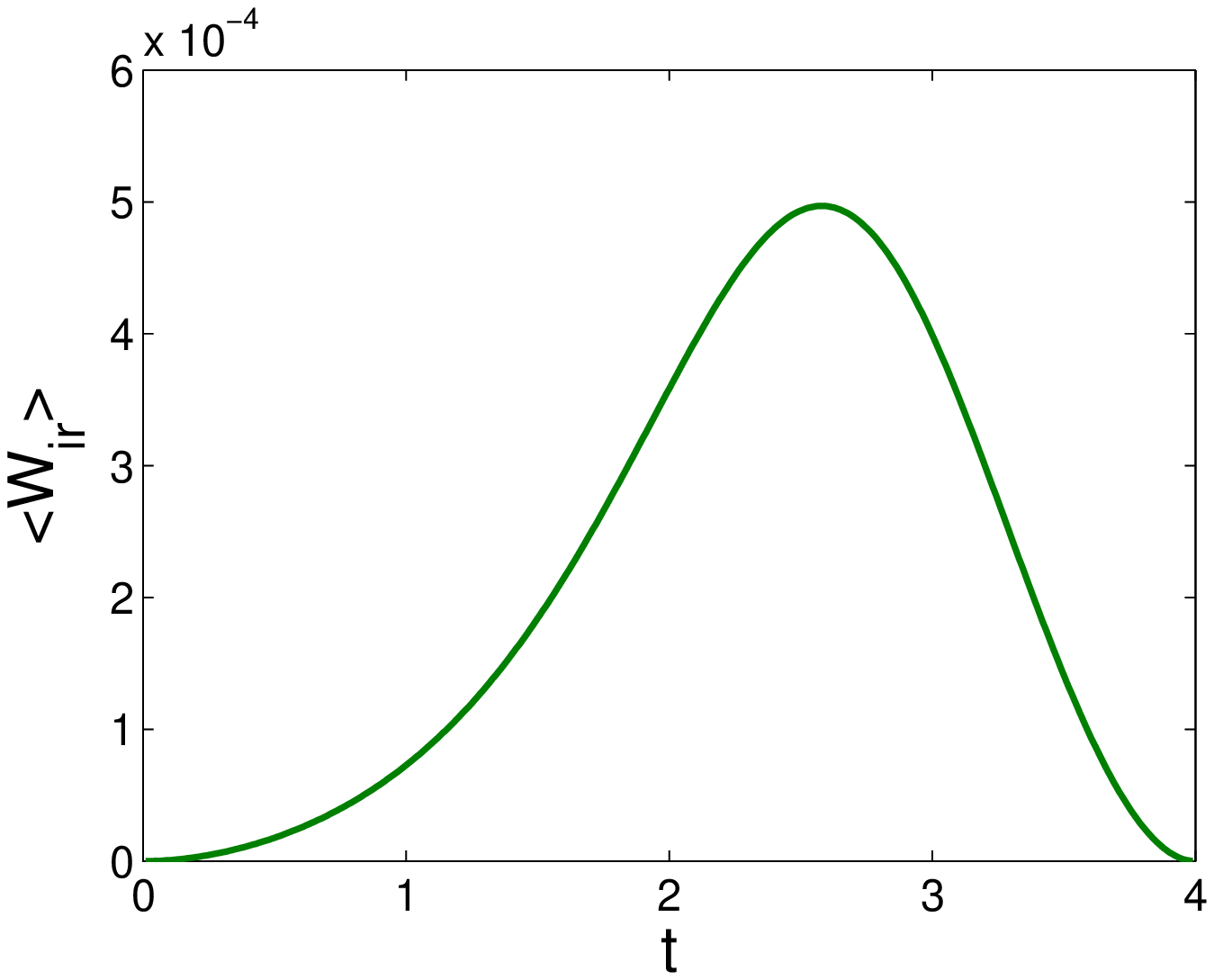}
	\caption{\label{irWork} $\big<W_\mathrm{ir}\big>$ for the values of the parameters $J_1=1, J_2=-1, B=0.1, d_0=2.5$. Adapted from \cite{Chotorlishvili}}
	\label{fig:irWork}
\end{figure}
\begin{figure}
	\centering
		\includegraphics[width=0.80\textwidth]{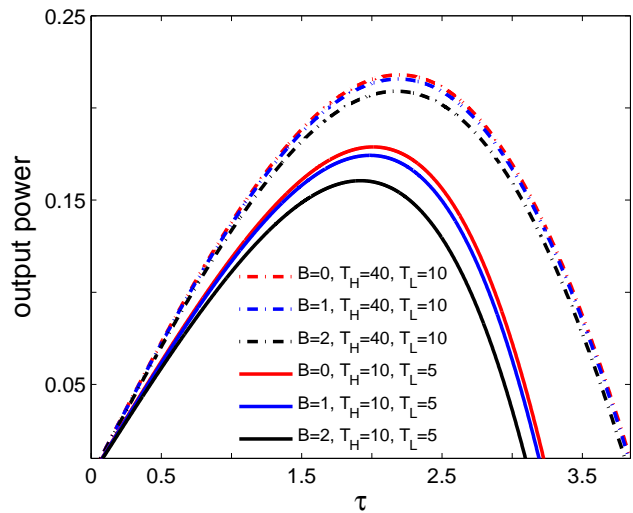}
	\caption{Output power for $J_1=1, J_2=-1, d_0=2.5$, $\varepsilon =1$ and  for the parameters depicted on the figure. Adapted from \cite{Chotorlishvili}}
	\label{fig:power}
\end{figure}

Shortcuts to adiabaticity allows  realizing  transitionless fast quantum adiabatic dynamics to  achieve a finite power output from a quantum engine.
We derived analytical results for the mean square fluctuation for the work, the irreversible work and output power of the cycle.
We observed that the work mean square fluctuations are increasing with the duration of the adiabatic strokes $\tau$  (cf. Fig.~\ref{rms}).
The irreversible work exhibits a non-monotonic behavior (see Fig.~\ref{fig:irWork}) and has a maximum for $\tau=0.26$(ps). At the end of the adiabatic stroke the irreversible work becomes zero confirming so that the cycle is reversible.
We found that the quantum heat engine with a MF working medium has an optimal duration corresponding to the largest power output (see Fig.~\ref{fig:power})

\begin{figure}
    \centering
      \includegraphics[width=0.80\textwidth]{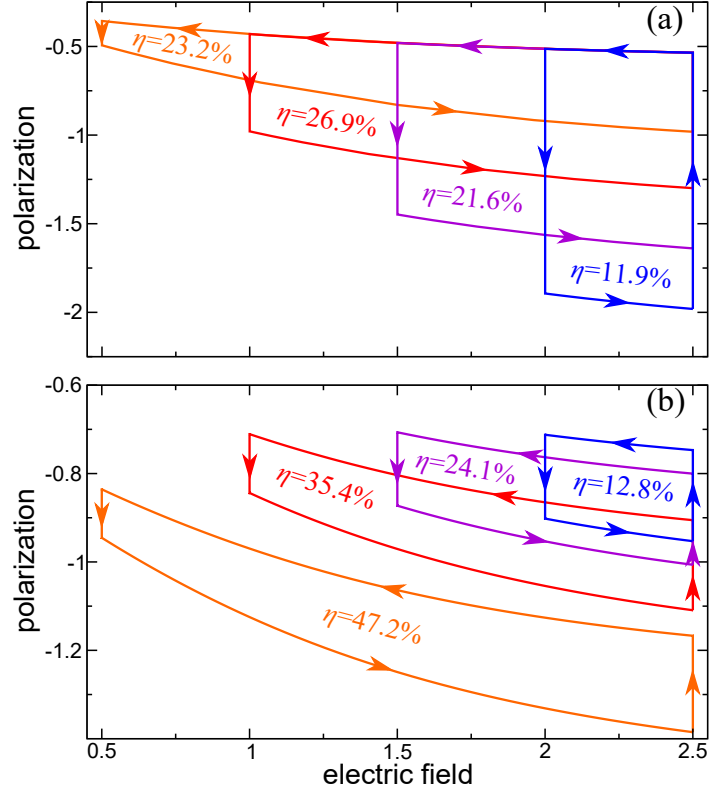}
    \caption{A complete quantum Otto cycle (a) using the level population corresponding to Gibbs distribution and (b) the level population obtained from the
     Lindblad master equation for the parameters $\gamma=0.1, T_H=40, T_L=10, d_0=2.5$ and $d_1$ as in the figures. Adapted from \cite{Chotorlishvili}}
    \label{cycle}
\end{figure}

By implementing a Lindblad master equation we studied the thermal relaxation of the system. We evaluated the transferred heat $\delta Q_{H}$ to the working substance   and the heat released by system to the cold bath $\delta Q_{c}$. We inferred  a cycle  efficiency of  $\eta = 1+\delta Q_{c}/\delta Q_{H}\approx 47\%$. If the
system thermalizes to the Gibbs ensemble the efficiency is lower  at $\eta \approx 23\%$  (see Fig.~\ref{cycle}).

\section*{Acknowledgments}
This work is supported by the DFG through the SFB 762.


\end{document}